# Slow Light Augmented Fabry-Perot Cavity for Enhanced Sensitivity in Measuring Frequency Shift

**Ruoxi Zhu[1*], Zifan Zhou[1], Dustin Greenwood[2], Jason Bonacum[3], David D. Smith[4], and Selim M. Shahriar[1,2]**

[1]*Department of Electrical and Computer Engineering, Northwestern University, Evanston, IL 60208, USA*

[2]*Department of Physics and Astronomy, Northwestern University, Evanston, IL 60208, USA*

[3]*Digital Optics Technologies, Rolling Meadows, IL 60008, USA*

[4]*NASA Marshall Space Flight Center*

[*]*Email: ruoxizhu2024@u.northwestern.edu*

**Abstract**

Recently, it has been shown that a slow-light augmented unbalanced Mach-Zehnder interferometer (SLAUMZI) can be used to enhance significantly the sensitivity of measuring the frequency shift of a laser, compared to the heterodyne technique. Here, we show that a similar enhancement can be realized using a slow-light augmented Fabry-Perot Cavity (SLAFPC), due to the fact that an FPC is inherently unbalanced, since different bounces of the field traverse different path lengths before interfering with the other bounces. We show how the degree of enhancement in sensitivity depends on the spectral width of the laser and the finesse of the FPC. For potentially realizable conditions, we show that a sensitivity enhancement factor ~$2.8*10^6$ can be achieved using a SLAFPC.

## 1. Introduction

Measuring the shift of the frequency of a laser can be used for many sensing applications. Examples include ring laser gyroscopes and accelerometers [1,2] and detectors for virialized ultra-light field dark matter [3]. To improve the sensitivity of such measurements, several schemes have been proposed employing superluminal lasers based on the fast-light effect [4,5,6,7,8,9,10,11,12,13,14,15,16]. However, it is also possible to achieve a sensitivity enhancement using the slow-light effect. Specifically, we have shown that a slow-light augmented unbalanced Mach-Zehnder interferometer (SLAUMZI) can be used to enhance significantly the sensitivity of measuring the shift in the frequency of a laser [17], compared to the heterodyne technique [2,18]. The idea is related to the fact that a slow-light medium inside an unbalanced Mach-Zehnder interferometer amplifies the number of fringes generated for a given change in the mean frequency, as shown in Ref. [19]. However, this work did not take into account the spectral width of the laser, and how that affects the minimum measurable frequency shift (MMFS). Here, we show that a similar enhancement can be realized using a slow-light augmented Fabry-Perot Cavity (SLAFPC), due to the fact that an FPC is inherently unbalanced, since different bounces of the field traverse different path lengths before interfering with the other bounces. We show how the degree of enhancement in sensitivity depends on the spectral width of the laser, the finesse of the FPC, and the loss caused by the medium that produces the slow-light process.

The rest of the paper is organized as follows. In Section 2, we first derive the transfer function of an empty ring cavity. Then we determine how the transfer function is modified in the presence of an intra-cavity slow-light medium, under the condition that the laser spectrum is a delta function. In Section 3, we show how the transfer function is modified further when we take into account the effect of the finite spectral width of the laser [20,21,22]. In Section 4, we first derive the expression for the MMFS if the SLAFPC is used for the measurement. Then we compare this with the MMFS for the heterodyne technique, to determine the sensitivity enhancement factor (SEF). In Section 5, we show how the loss due to the presence of the slow-light medium may affect the sensitivity of the SLAFPC. Concluding remarks are presented in Section 6. The Supplement shows the details involved in deriving the signal for an FPC when taking into account the effect of the finite width of the laser spectrum.

## 2. Model for a Fabry-Perot Cavity Driven by an Ideal Laser with a Delta Function Linewidth

### *a. Empty cavity case with ideal laser input:*

To derive the transfer function of a Fabry-Perot cavity, we model it as a ring resonator, as shown in Figure 1. The laser entering the cavity is initially assumed to have a delta function spectral width. The cavity is composed of three legs, each with a physical length $l$, so that the cavity length is $L = 3l$. The input and output couplers are assumed to be identical, with a real reflection and transmission coefficients denoted as $r$ and $t$, respectively, while the other mirror is a perfect reflector. An anti-reflection coating is assumed to have been applied to the intra-cavity side of the two couplers. The thickness of each coupler is assumed to be negligible. A π phase shift is added for reflection at the air-to-glass interface [23]. Assuming there is no loss inside the cavity, we get $R=r^2=1-T=1-t^2$, where $R$ ($T$) is the intensity reflectivity (transmittivity). Therefore, in steady state, and using the notation illustrated in Figure 1, we can write the electric fields at various ports as follows:

$$E_r = -rE_i + E_B^a t;\ E_o = E_F^b t;\ E_F^b = E_F^a e^{jkl};\ E_B^b = rE_F^b;\ E_B^a = E_F^b e^{j2kl};\ E_F^a = tE_i + rE_B^a \tag{1}$$

where $k = 2\pi/\lambda$ is the wave number and $\lambda$ is the vacuum wavelength of the laser.

Solving the equations above, the ratio of the output field $E_o$ and the input field $E_i$ can be expressed as [24]:

$$\frac{E_o}{E_i} = \frac{t^2 e^{jkl}}{1 - r^2 e^{jkL}} = \frac{Te^{jkl}}{1 - Re^{jkL}} \tag{2}$$

The transfer function of the cavity is defined as:

$$H \equiv \left|\frac{E_o}{E_i}\right|^2 = \frac{T^2}{\left|1 - Re^{jkL}\right|^2} = \frac{1}{1 + \left(\frac{4R}{T^2}\right)\sin^2\left(\frac{kL}{2}\right)} = \frac{1}{1 + \left(\frac{4R}{T^2}\right)\sin^2\left(\frac{\pi f L}{c_0}\right)} \tag{3}$$

where $f$ is the frequency of the laser in Hz and $c_0$ is the speed of light in vacuum. Eqn. (3) is well known, but is derived here in order to establish clearly the assumption and the notations. The empty-cavity free spectral range in terms of the angular frequency is $FSR|_{\omega}^{EC} = 2\pi c_0 / L$.

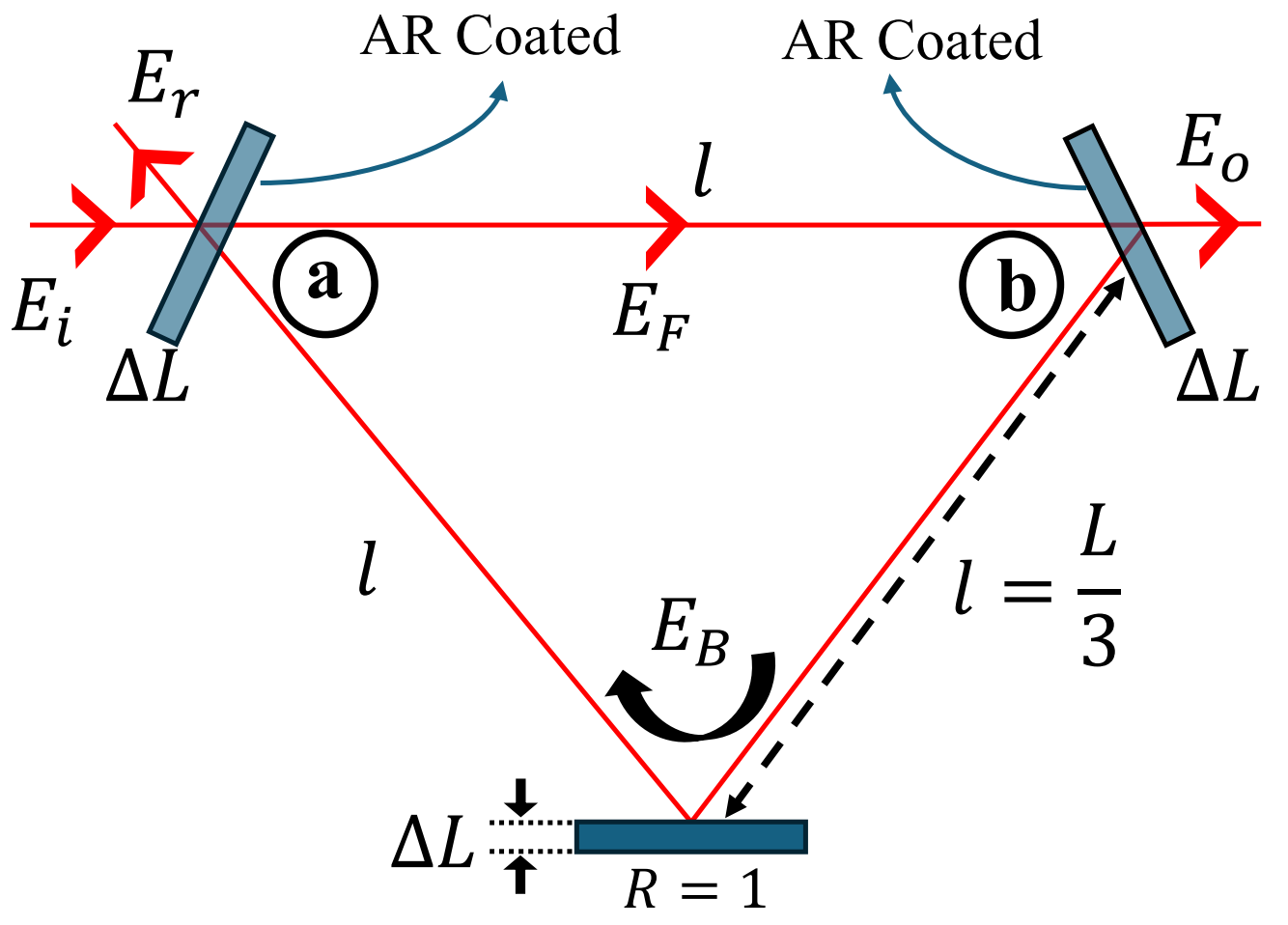

Figure 1. Schematic illustration of an empty Fabry-Perot resonator. The thickness $\Delta L$ of each reflector is assumed to be negligible.

To obtain the linewidth of the transfer function, we assume that there is a resonance frequency of $\omega_m$ and define a deviation from the resonance frequency as $\tilde{\omega} = \omega - \omega_m$, which is much less than a free spectral range. Therefore, the transfer function at frequency $\omega$ can be written as:

$$H(\omega) = \frac{1}{1 + \frac{4R}{T^2}\sin^2\left(\frac{\omega L}{2c_0}\right)} = \frac{1}{1 + \frac{4R}{T^2}\sin^2\left(\frac{(\tilde{\omega} + \omega_m)L}{2c_0}\right)} \approx \frac{1}{1 + \frac{R}{T^2}\frac{\tilde{\omega}^2 L^2}{c_0^2}} = \frac{\gamma_{EC}^2}{\gamma_{EC}^2 + \tilde{\omega}^2} \tag{4}$$

Here, we have made the approximation that $\tilde{\omega} L / c_0 \ll 1$ and $\gamma_{EC} \equiv c_0 T / L\sqrt{R}$ is defined as the halfwidth at half maximum of the transfer function (HWHM). We define $\tau_{EC} = \gamma_{EC}^{-1}$ as the empty cavity decay time. The full width at half maximum (FWHM) $\tilde{\gamma}$ and the finesse $F$ of the cavity can be expressed as:

$$\tilde{\gamma} = 2\gamma_{EC};\; F = \frac{FSR|_{\omega}^{EC}}{\tilde{\gamma}} = \frac{\pi\sqrt{R}}{T} = \frac{\pi\sqrt{R}}{1-R} \tag{5}$$

We define $\tau_{EC} = \gamma_{EC}^{-1}$ as the empty cavity decay time, $\Delta\phi_{EC} = \tilde{\omega}\tau_{EC}$ as the effective dephasing parameter for the empty cavity. In terms of these, the transfer function can be expressed as:

$$H(\omega) = \frac{1}{1 + (\tilde{\omega}\tau_{EC})^2} = \frac{1}{1 + \Delta\phi_{EC}^2} \tag{6}$$

### *b. Slow light case with ideal laser input*

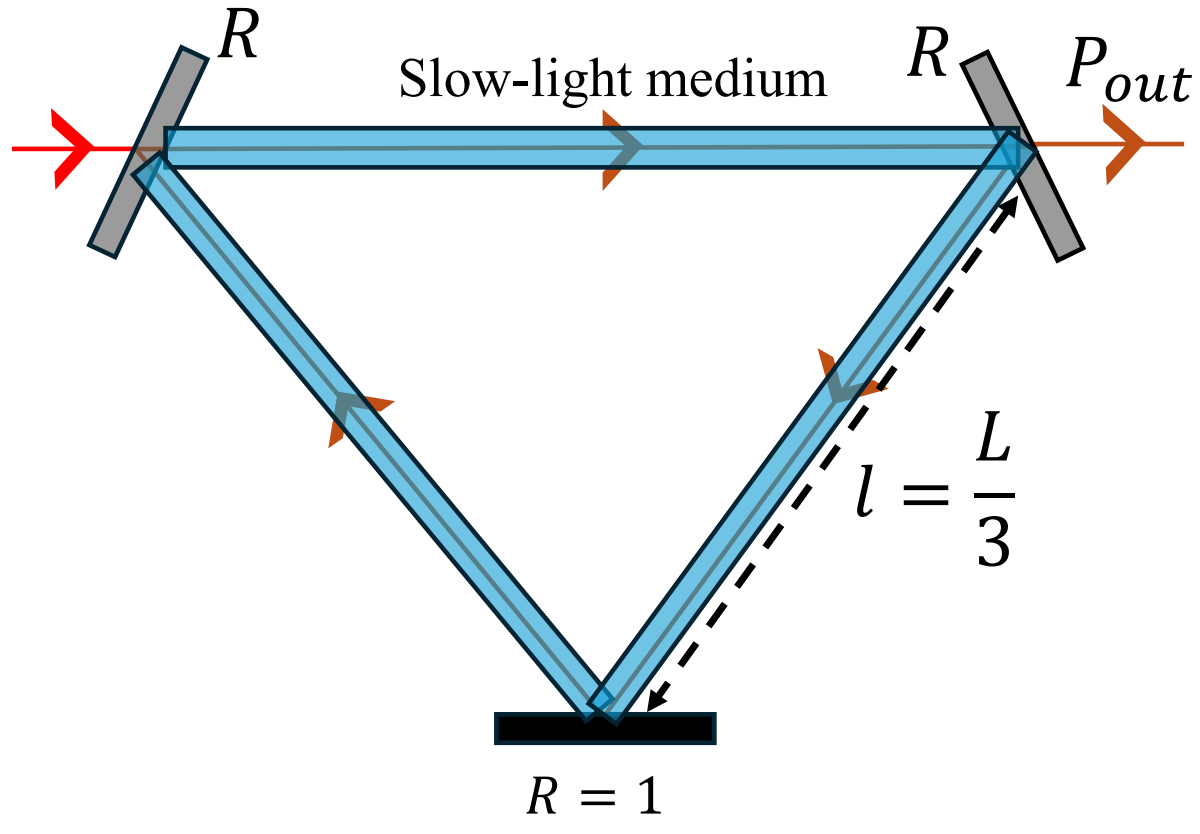

Figure 2. Schematic illustration of a slow-light augmented Fabry-Perot cavity (FPC).

In the presence of a slow-light medium in the cavity, as shown in Figure 2, the phase accumulation through propagation is multiplied by the refractive index of the medium. The case when the cavity is only partially filled with a slow-light medium can be analyzed via a simple modification of the following analysis. To the first order, the refractive index is a linear function of frequency variation. The linear approximation is valid when the frequency only varies within a small range around the center frequency corresponding to the unity index, which is the case in high-finesse cavities. Therefore, without loss of generality, it is convenient to set the center frequency at the *m-th* resonance frequency of the empty cavity, $\omega_m$, so that $n(\omega) \simeq 1+\sigma(\omega-\omega_m) \simeq 1+\sigma\tilde{\omega}$, where $\sigma$ is a coefficient whose value depends on the particular medium used. The transfer function is then modified to be:

$$H(\omega) = \frac{1}{1+\frac{4R}{T^2}\sin^2\left(\frac{n(\omega)*\omega L}{2c_0}\right)} = \frac{1}{1+\frac{4R}{T^2}\sin^2\left(\frac{(1+\sigma\tilde{\omega})(\omega_m+\tilde{\omega})L}{2c_0}\right)} \tag{7}$$

Ignoring the $\tilde{\omega}^2$ term inside the sine function and making use of $\omega_m L/2c_0 = m\pi$, the transfer function can be simplified to:

$$H(\tilde{\omega}) = \frac{1}{1+\frac{2R}{T^2}\left[1-\cos\left(\frac{\tilde{\omega}L}{c_0}n_g\right)\right]} \approx \frac{\gamma_{EC}^2}{\gamma_{EC}^2+\left(\tilde{\omega}n_g\right)^2} \tag{8}$$

where $n_g = 1+\sigma\omega_m$ is the group index experienced by the laser and $\gamma_{EC}$ is the half width at half maximum (HWHM). Similar to the empty cavity case, here we have made the approximation that $\tilde{\omega}Ln_g / c_0 \ll 1$. The free spectral range in the slow-light case is obtained by making use of Eqn. (8) and is determined to be $FSR\big|_{\omega}^{SL} = 2\pi c_0/\left(n_g L\right) = FSR\big|_{\omega}^{EC}/n_g$. By defining the slow light case HWHM as $\gamma_{SL} = \gamma_{EC}/n_g$, the transfer function can be written as:

$$H(\tilde{\omega}) = \frac{\gamma_{SL}^2}{\gamma_{SL}^2 + \tilde{\omega}^2} \tag{9}$$

We define $\Delta\phi_{SL} = \Delta\phi_{EC} n_g = \tilde{\omega}\tau_{EC} n_g$ as the effective dephasing parameter in the presence of a slow-light medium in the cavity. In terms of this parameter, we can write the transfer function as:

$$H(\tilde{\omega}) = \frac{1}{1+\left(\tilde{\omega}\tau_{EC} n_g\right)^2} = \frac{1}{1+\Delta\phi_{SL}^2} \tag{10}$$

It should be noted that these expressions, while derived for slow-light, is valid for both positive (slow-light) and negative (fast-light) values of $n_g$. For the case when the group index is negative and has a magnitude much less than unity, this corresponds to the well-known case of the so-called white light cavity [25]

## 3. Accounting for the effect of the finite laser linewidth

In order to determine the enhancement in sensitivity achievable with the SLAFPC, it is essential to take into account the finite spectral width of the laser. In the quantum limit, this is due to the addition of the vacuum mode, and the spectrum is Lorentzian, with a width given by the so-called Schwalow-Townes Linewidth (STL). In some cases, the linewidth can be broader than the STL, due to other effects that cause random phase jumps [26,27,28]. In what follows, we restrict our discussion to the case where the linewidth is the STL.

We express the output field as a summation of the electric field in each round trip inside the cavity:

$$E_0 = E_i T e^{jkl} \sum_{n=0}^{\infty} \varepsilon_n = E_i T e^{jkl} \sum_{n=0}^{\infty} R^n e^{jnkL} e^{j\phi_n}. \tag{11}$$

Here, the index *n* indicates the *n-th* round trip inside the cavity and $\phi_n$ is an additional phase factor that accounts for the random phase jumps that cause the spectral width of the laser. We define $\Delta\phi_{nm} \equiv \left(\phi_n - \phi_m\right)$ to represent the random phase difference between the fields in the *n-th* and *m-th* round trips. Under the assumption that the spectral width of the laser is Lorentzian, the effect of which can be modeled as random walk [29], this phase factor has a Gaussian probability distribution:

$$P\left(\Delta\phi_{nm}\right) = \frac{1}{\sqrt{2\pi\tau_{nm}/\tau_{STL}}} \exp\left[-\frac{\Delta\phi_{nm}^2}{2\left(\tau_{nm}/\tau_{STL}\right)}\right] \tag{12}$$

where

$$\tau_{nm} = \left|n-m\right|\tau_{delay} \tag{13}$$

Here, $\tau_{STL} = \gamma_{STL}^{-1}$ is the coherence time of the laser, $\gamma_{STL}$ is the STL, and $\tau_{delay}$ is the time delay for a round trip. Regardless of the presence of the slow-light medium, the time delay can be expressed as the round trip time in vacuum ($\tau_{RT}$), i.e., $\tau_{delay} = \tau_{RT}$, for the following reasons. First, since the spectral width of the laser is due strictly to the STL caused by spontaneous emission, the phase jumps can be modeled as occurring instantaneously (according to the Markov approximation used in modeling the coupling between the atoms and the bath of vacuum modes), corresponding to a bandwidth that approaches infinity. On the other hand, the bandwidth of the positive dispersion that causes the slow-light effect is typically very small (of the order of a few MHz). As such, it is a good approximation to assume that the phase jumps would propagate through the slow-light medium at essentially the phase velocity rather than the group velocity. Second, in the case where a gas vapor is employed to generate the slow-light effect, the refractive index of the diluted vapor can be well approximated as unity; thus, the phase velocity equals the speed of light in vacuum. It then can be shown [29] that when averaged over such a probability distribution, we get:

$$\left\langle \sin\left(\Delta\phi_{nm}\right)\right\rangle = 0; \quad \left\langle \cos\left(\Delta\phi_{nm}\right)\right\rangle = e^{-\tau_{nm}/2\tau_{STL}} \tag{14}$$

We can thus write:

$$\left\langle e^{-i\Delta\phi_{nm}} \right\rangle = e^{-\tau_{delay}|m-n|/2\tau_{STL}} \tag{15}$$

In steady state, the transfer function can be expressed as:

$$H = T^2 \sum_{n=0}^{\infty}\sum_{m=0}^{\infty} \left\langle \varepsilon_n \varepsilon_m^* \right\rangle \tag{16}$$

In the supplement, we show in detail that the double summation can be turned into a single summation and the obtained result agrees with the one in Ref. [30]. Here, we only give a brief derivation for simplicity. Considering that the interference between *n-th* and *m-th* round trip is attenuated by a factor of $e^{-\tau_{delay}|m-n|/2\tau_{STL}}$, we can define $Q \equiv \tau_{delay}/2\tau_{STL}$ as a measure of laser decoherence within one round trip of the cavity and rewrite $H$ as:

$$H = T^2 \left( \sum_{n=0}^{\infty} \left|\varepsilon_n\right|^2 + \sum_{\substack{n=0 \\ n\neq m}}^{\infty} \sum_{m=0}^{\infty} \varepsilon_n \varepsilon_m^* e^{-Q|m-n|} \right) \tag{17}$$

where on the right-hand side, the single summation is equivalent to $1/\left(1-R^2\right)$ and the double summation can be calculated to be:

$$\begin{aligned} \sum_{\substack{n=0 \\ n\neq m}}^{\infty} \sum_{m=0}^{\infty} \varepsilon_n \varepsilon_m^* e^{-Q|m-n|} &= \sum_{s=0}^{\infty}\sum_{q=1}^{\infty} \left( \varepsilon_s \varepsilon_{s+q}^* + \varepsilon_{s+q} \varepsilon_s^* \right) e^{-Q|q|} \\ &= \left( \sum_{n'=-\infty}^{+\infty} \left(Re^{-Q}\right)^{|n'|} e^{jn'kL} - 1 \right) \sum_{s=0}^{\infty} R^{2s} = \frac{1}{1-R^2} \left( \sum_{n'=-\infty}^{+\infty} \left(Re^{-Q}\right)^{n'} e^{jn'kL} - 1 \right) \end{aligned} \tag{18}$$

Therefore, the transfer function can be expressed as [30]:

$$H = \frac{a(\Delta)}{1 + b(\Delta)\sin^2\left(\frac{\omega L}{2c_0}\right)} \tag{19}$$

where

$$a(\Delta) = \frac{T^2\left(1-\Delta^2 R^2\right)}{\left(1-R^2\right)\left(1-\Delta R\right)^2}; \quad b(\Delta) = \frac{4\Delta R}{\left(1-\Delta R\right)^2}; \quad \Delta \equiv e^{-Q} \tag{20}$$

As before, we set the *m-th* resonance frequency at $\omega_m$ and define $\tilde{\omega} \equiv \omega - \omega_m$, which is again assumed to be much less than an FSR. Consider first the case where a slow-light medium is not present. By expanding around $\omega_m$ and making the approximation that $\tilde{\omega}L/c_0 \ll 1$, we get:

$$\tilde{H}(\tilde{\omega}) = \frac{a(\Delta)\tilde{\gamma}_{EC}^2}{\tilde{\gamma}_{EC}^2 + \tilde{\omega}^2} \tag{21}$$

where $\tilde{\gamma}_{EC} = 2c_0 / L\sqrt{b(\Delta)}$ is the HWHM, incorporating the effect of the laser linewidth in the empty cavity case. It should be noted that in the limit where an ideal laser with delta function linewidth is used, $\Delta = 1$, so that $a(\Delta) = a(0) = 1$ and $b(\Delta) = b(0) = 4R/T^2$, recovering Eqn. (4)

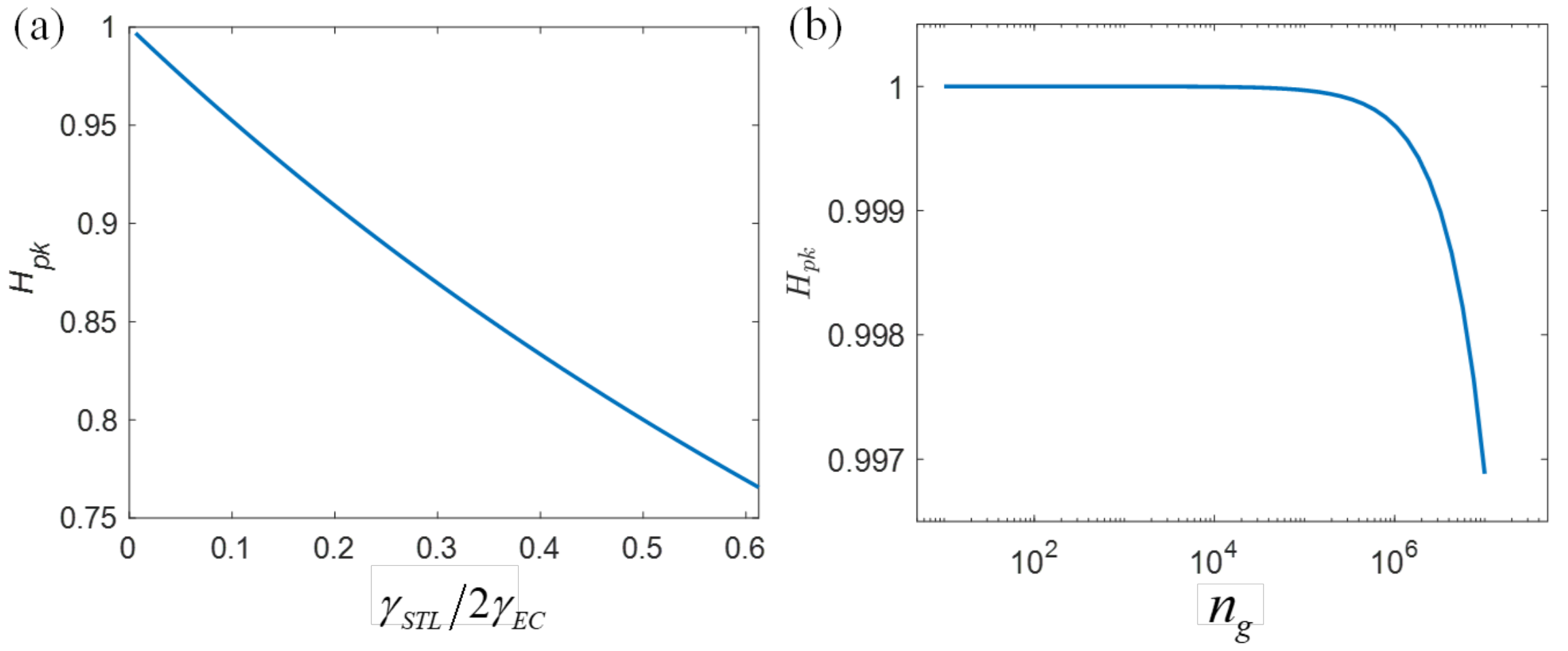

Figure 3. (a). Peak transmission as a function of $\gamma_{STL}$ at unity group index. (b). Peak transmission with respect to the group index, for $\gamma_{STL}/2\gamma_{EC} \simeq 0.012$.

. However, when the STL is finite, $\Delta < 1$, so that $a(\Delta) < a(0)$ and $b(\Delta) < b(0)$. This corresponds to a reduction of the peak amplitude of transmission, and a corresponding broadening of the FPC transmission spectrum. It can be shown that if $\gamma_{EC} \ll \gamma_{STL}$, then the value of $\tilde{\gamma}_{EC}$ approaches $\gamma_{STL}$. This is as expected, since the transmission spectrum cannot be narrower than the STL.

Consider next the case where the cavity is filled with a slow-light medium. Comparing Eqn. (21) with Eqn. (8), the effect of slow light for linearly varying index around $\omega_m$ is obtained by substituting $\tilde{\omega}$ with $n_g\tilde{\omega}$:

$$\tilde{H}_{SL}(\tilde{\omega}) = \frac{a(\Delta)\tilde{\gamma}_{EC}^2}{\tilde{\gamma}_{EC}^2 + (\tilde{\omega}n_g)^2} = a(\Delta)\cdot\frac{\tilde{\gamma}_{SL}^2}{\tilde{\gamma}_{SL}^2 + \tilde{\omega}^2}; \quad \Delta = e^{-Q} \tag{22}$$

where $\tilde{\gamma}_{SL} = \tilde{\gamma}_{EC}/n_g$ and $Q = \tau_{delay}/2\tau_{STL} = \tau_{RT}/2\tau_{STL}$. Thus, the only effect of the slow-light medium is to reduce the spectral width of the FPC transmission, by the factor of the group index. Of course, we have not included here the deleterious effect of possible losses introduced by the slow-light medium. We address this issue in Section 5.

To illustrate the effect of a finite laser linewidth, consider an FPC for which the reflectivity *R* is 0.95, and the cavity length is 0.3 meters. The HWHM for the empty cavity $\gamma_{EC}$ for an ideal laser ($\tau_{STL} \rightarrow \infty$) for such an FPC is ~5.1*$10^7$ $sec^{-1}$. Figure 3 plots the peak transmission as a function of both $\gamma_c = \gamma_{STL}/2$ and $n_g$.

## 4. Minimum Measurable Frequency Shift

As presented in Ref. [31], the minimum measurable frequency shift (MMFS) is determined by a combination of two effects: the shot-noise due to the vacuum mode (VM) and the phase fluctuation caused by the spontaneous emission (SE). Therefore, we first break down the MMFS analysis into parts corresponding to the VM and the SE separately, and then present the overall MMFS.

### *a. MMFS Due to the Vacuum Mode*

Consider a situation where we are interested in measuring the shift in the frequency of a test laser. To measure the shift, we assume that the whole output of the laser is sent through the SLAFPC, and the laser output is defined as $S_0 = N\tau_s$, where *N* is the number of emitted photons per second and $\tau_S = \gamma_{EC}^{-1}$ is the time for a single measurement, so that the bandwidth of the single measurement is $\gamma_S\tau_S^{-1} = \gamma_{EC}$. Then the output signal from the SLAFPC can be written as:

$$S = S_0\tilde{H}(\tilde{\omega}) = S_0\cdot a(\Delta)\frac{\tilde{\gamma}_{SL}^2}{\tilde{\gamma}_{SL}^2 + \tilde{\omega}^2} \tag{23}$$

We now consider using the hopping technique to measure the MMFS. To implement such a technique, the laser frequency is assumed to be in close proximity of $\tilde{\omega}_0 = 0$, corresponding to the maximum transmission, and is modulated by a square function, which makes the laser frequency swing between $\tilde{\omega} + \kappa\tilde{\gamma}_{SL}$ and $\tilde{\omega} - \kappa\tilde{\gamma}_{SL}$, where $\kappa$ is a parameter that determines the amplitude of the swing and needs to be optimized. The error signal is then defined as $S_H$ and is determined by the difference of the output signal at $\tilde{\omega} + \kappa\tilde{\gamma}_{SL}$ and $\tilde{\omega} - \kappa\tilde{\gamma}_{SL}$:

$$S_H(\tilde{\omega},\kappa) \equiv S_R - S_L = S(\tilde{\omega}+\kappa\tilde{\gamma}_{SL}) - S(\tilde{\omega}-\kappa\tilde{\gamma}_{SL}) \tag{24}$$

where $S_{R(L)}$ is the output signal at $\tilde{\omega}+\kappa\tilde{\gamma}_{SL}$ ($\tilde{\omega}-\kappa\tilde{\gamma}_{SL}$).

Following the definition in Ref. [31], the minimum measurable frequency shift due to the vacuum mode is written as:

$$\Delta\omega_{SLAFPC,VM} = \frac{\Delta S_H}{\left|\partial S_H / \partial\tilde{\omega}\right|} \tag{25}$$

Assuming the fluctuation in the signal, $\Delta S_H$, to be due to shot noise [32] and minimizing the MMFS with respect to $\kappa$, the value of $\kappa$ is determined to be $1/\sqrt{2}$. Therefore, we can write:

$$\begin{aligned} \Delta S\Big|_{\tilde{\omega}=\tilde{\gamma}_{SL}/\sqrt{2}} &= \sqrt{\left(\Delta S_R\right)\Big|_{\tilde{\omega}=\tilde{\gamma}_{SL}/\sqrt{2}}^{2} + \left(\Delta S_L\right)\Big|_{\tilde{\omega}=\tilde{\gamma}_{SL}/\sqrt{2}}^{2}} = \sqrt{\frac{4}{3}S_0 a(\Delta)} \\ \frac{\partial S}{\partial\tilde{\omega}}\Big|_{\tilde{\omega}=\tilde{\gamma}_{SL}/\sqrt{2}} &= \frac{\partial S_R}{\partial\tilde{\omega}}\Big|_{\tilde{\omega}=\tilde{\gamma}_{SL}/\sqrt{2}} - \frac{\partial S_L}{\partial\tilde{\omega}}\Big|_{\tilde{\omega}=\tilde{\gamma}_{SL}/\sqrt{2}} = \frac{8\sqrt{2}}{9}S_0 a(\Delta)\frac{1}{\tilde{\gamma}_{SL}} \end{aligned} \tag{26}$$

Thus, the MMFS in the SLAFPC is:

$$\Delta\omega_{SLAFPC,VM} = \frac{3\sqrt{6}}{8}\frac{\tilde{\gamma}_{SL}}{\sqrt{S_0 a(\Delta)}} \equiv \beta\cdot\frac{\tilde{\gamma}_{EC}}{n_g\sqrt{2S_0 a(\Delta)}} \tag{27}$$

where $\beta \simeq 1.3$.

We can express $\tilde{\gamma}_{EC}$ in terms of $\gamma_{EC}$ by using Eqn. (21):

$$\tilde{\gamma}_{EC} = \frac{2c_0}{L}\cdot\frac{1}{b(\Delta)} = \frac{c_0 T}{LR}\cdot\sqrt{\frac{b(\Delta=1)}{b(\Delta)}} = \gamma_{EC}\cdot\frac{1-R\Delta}{(1-R)\Delta^{1/2}} \equiv \gamma_{EC}\cdot\alpha \tag{28}$$

Therefore, the MMFS for SLAFPC can be simplified to:

$$\Delta\omega_{SLAFPC,VM} = \beta\sqrt{\frac{1}{2S_0}}\cdot\frac{\gamma_{EC}}{n_g}\cdot\sqrt{\frac{\alpha^2}{a(\Delta)}} \equiv \beta\sqrt{\frac{1}{2S_0}}\cdot\frac{\gamma_{EC}}{n_g}\cdot\xi \tag{29}$$

where

$$\xi = \frac{(1-R\Delta)^2}{(1-R)^2}\cdot\sqrt{\frac{1-R^2}{\Delta(1-\Delta^2R^2)}} = \frac{(1-R\Delta)^2}{T^2}\cdot\sqrt{\frac{1-R^2}{\Delta(1-\Delta^2R^2)}} \tag{30}$$

Utilizing $S_0 = N\tau_S = \gamma_{LC}^2\,\tau_S/2\gamma_{STL}$, where $\gamma_{LC}$ is the cavity linewidth of the input laser, and noting that $\gamma_S = \gamma_{EC}$ we can express $\Delta\omega_{SLAFPC,VM}$ as:

$$\Delta\omega_{SLAFPC,VM} = \frac{1}{n_g}\beta\xi\frac{\gamma_S}{\gamma_{LC}}\cdot\sqrt{\gamma_S\gamma_{STL}}\,. \tag{31}$$

If the measurement is repeated $M = \tau_M / \tau_S = \gamma_S / \gamma_M$ times, where $\tau_S$ is the total time of measurement and $\gamma_M$ is the measurement bandwidth, the frequency uncertainty is reduced by a factor of $\sqrt{M}$ , as we get:

$$\Delta\omega_{SLAFPC,VM} = \frac{1}{n_g}\beta\xi\frac{\gamma_S}{\gamma_{LC}}\cdot\sqrt{\gamma_M\gamma_{STL}} = \frac{1}{n_g}\beta\xi\varepsilon_0\Gamma_0 . \quad (32)$$

where $\Gamma_0 \equiv (\gamma_M\gamma_{STL})^{1/2}$ and $\varepsilon_0 \equiv \gamma_S/\gamma_{LC} = \gamma_{EC}/\gamma_{LC}$ is parameter that represents the ratio of the linewidths of the empty FPC and the laser cavity.

For a typical laser of metrological interest, the STL is significantly smaller than 1 sec$^{-1}$. On the other hand, the cavity decay rate for a typical FPC, even with a very high finesse, is much larger than 1 sec$^{-1}$. As such, the operating condition would correspond to $\tau_{EC} = \tau_S \ll \tau_{STL}$. In this limit, we have $\Delta \to 1$, which in turn corresponds to $\xi \to 1$. In this limit, noting that $\beta$ is close to unity, we get:

$$\Delta\omega_{SLAFPC,VM} \simeq \frac{\varepsilon_0}{n_g}\Gamma_0 . \quad (33)$$

Thus, in order to make the VM contribution to the MMFS negligible compared to $\Gamma_0$, the group index must be significantly larger than $\varepsilon_0$, which is the ratio of the linewidths of the empty FPC and the laser cavity.

It is also instructive to consider the special case where $n_g = 1$, corresponding to the Fabry-Perot case derived in Section 3.2 in Ref. [31]. For this case, it is shown [31] that the global minimum for the vacuum mode contribution to the MMFS for the FPC occurs at $\tau_S = \tau_{EC} = 4\tau_{STL}$ . The corresponding value of the MMFS due to the VM is ~ $0.85*\Gamma_0/\langle n\rangle_{LC}$ , where $\langle n\rangle_{LC}$ is the photon number in the laser cavity, which is very large for most lasers of interest. As such, the MMFS due to the VM can be made negligibly small compared to $\Gamma_0$ in principle even without using slow-light. However, the condition for reaching this value, namely $\tau_S = \tau_{EC} = 4\tau_{STL}$ , means that the product of the finesse and the total length of the FPC must be ~4 times larger than the coherence length of the laser. Consider, for example, a laser with an STL of 1 sec$^{-1}$, so that the coherence length is ~3x10$^8$ meters. For a state of the art FPC with a finesse of ~3x10$^5$, this would require the cavity length to be ~4 km, which is obviously impractical. On the other hand, use of slow-light with a group index much larger than $\varepsilon_0$ would make the MMFS due to the VM negligible compared to $\Gamma_0$ without requiring such a large cavity length.

### *b. MMFS Due to Spontaneous Emission*

In Ref. [31], we have shown in details the general approach for determining the MMFS due to the effect of spontaneous emission (SE) into the lasing mode inside the laser cavity. To summarize briefly, the SE causes a phase diffusion, with a standard deviation of $\Delta\phi_{SE} = \sqrt{\tau/\tau_{STL}}$ , over an observation time of $\tau$. In the case of a conventional FPC, for a single measurement, we have $\tau = \tau_S = \tau_{EC}$. Since $\Delta\phi = \tilde{\omega}\tau_{EC} = \tilde{\omega}\tau_S$, as shown in Eqn. (6), it follows that for an FPC without a slow-light medium, the corresponding frequency uncertainty is given by:

$$\Delta\omega_{FPC,SE} = \Delta\phi_{SE}/\tau_S = \sqrt{1/(\tau_S\tau_{STL})} \tag{34}$$

If the measurement is repeated $M = \tau_M/\tau_S = \gamma_S/\gamma_M$ times, where $\tau_S$ is the total time of measurement and $\gamma_M$ is the measurement bandwidth, the frequency uncertainty is reduced by a factor of $\sqrt{M}$, and we get:

$$\Delta\omega_{FPC,SE} = \sqrt{1/(\tau_M\tau_{STL})} = \sqrt{\gamma_M\gamma_{STL}} = \Gamma_0 \tag{35}$$

When the FPC is filled with a slow-light medium, it is still the case that, for a single measurement, $\tau = \tau_S = \tau_{EC}$, since the phase jumps caused by the SE does not get slowed down due to the fact that the jumps are effectively instantaneous, while the slowing effect has a very narrow bandwidth, as discussed in Subsection 4.a above. Therefore, for the SLAFPC, we have $\Delta\phi_{SE} = \sqrt{\tau_S/\tau_{STL}}$. At the same time, any phase change in a SLAFPC can be expressed as $\Delta\phi = n_g\tilde{\omega}\tau_M$ in the presence of slow light, as shown in Eqn. (10). Therefore, the corresponding frequency uncertainty is given by:

$$\Delta\omega_{SLAFPC,SE} = \frac{\Delta\phi_{MIN}}{n_g\tau_S} = \frac{1}{n_g}\sqrt{\frac{1}{\tau_S\tau_{STL}}} \tag{36}$$

If the measurement is repeated $M = \tau_M/\tau_S = \gamma_S/\gamma_M$ times, where $\tau_S$ is the total time of measurement and $\gamma_M$ is the measurement bandwidth, the frequency uncertainty is reduced by a factor of $\sqrt{M}$. As such, the MMFS due to the SE for the SLAFPC is found to be:

$$\Delta\omega_{SLAFPC,SE} = \frac{1}{n_g}\sqrt{\frac{1}{\tau_M\tau_{STL}}} = \frac{\Gamma_0}{n_g} \tag{37}$$

### c. *Overall MMFS*

As we have shown in Ref. [31], the overall MMFS is given by the RMS value of the MMFS for due to the VM and the MMFS due to the SE. As such, for the SLAPFC, the overall MMFS can be expressed as:

$$\Delta\omega_{SLAFPC,tot} = \sqrt{\Delta\omega^2_{SLAFPC,SE} + \Delta\omega^2_{SLAFPC,VM}} = \frac{\Gamma_0}{n_g}\sqrt{1+(\beta\xi\varepsilon_0)^2}; \quad \varepsilon_0 = \frac{\gamma_S}{\gamma_{LC}} = \frac{\gamma_{EC}}{\gamma_{LC}} \tag{38}$$

In Ref. [31], we have shown that for conventional sensors one can in principle find operating conditions under which the MMFS due to the VM can be made much smaller than the MMFS due to the SE, so the overall MMFS has a lower bound of $\Gamma_0$, which is also considered to be the standard limit of the MMFS. Therefore, the sensitivity enhancement factor (SEF) for the SLAPFC can be defined as:

$$SEF = \frac{\Gamma_0}{\Delta\omega_{SLAFPC,tot}} = \frac{n_g}{\sqrt{1+(\beta\xi\varepsilon_0)^2}} \tag{39}$$

As noted earlier, for a typical laser of metrological interest, the STL is significantly smaller than 1 sec$^{-1}$. On the other hand, the cavity decay rate for a typical FPC, even with a very high finesse, is much larger than 1 sec$^{-1}$. As such, the operating condition would correspond to $\tau_{EC} = \tau_S \ll \tau_{STL}$. In this limit, we have $\Delta \to 1$, which in turn corresponds to $\xi \to 1$. In this limit, noting that $\beta$ is close to unity, we get:

$$SEF \simeq \frac{n_g}{\sqrt{1+\varepsilon_0^2}} \tag{40}$$

As an example, consider a test laser generated in a ring cavity at 632 nanometers, with a partial reflector whose intensity reflectivity is 0.9, and two perfect reflectors. The laser cavity is assumed to be 0.7 meters long and the output power is 10 mW. Therefore, the rate of number of photons incident on the input coupler of the SLAFPC is 3.2*10$^{16}$ sec$^{-1}$ , and the STL is calculated to be 0.032 sec$^{-1}$. Figure 4 shows the Sensitivity Enhancement Factor (SEF) as a function of the group index and the cavity finesse for the SLAFPC using such a laser, using the expression shown in Eqn. (39). For the maximum value (313) of the finesse shown in this plot, we have $\varepsilon_0 \simeq 0.22$. For these parameters, $\xi \simeq 1$. Given that $\beta \simeq 1.3$, the corresponding value of the SEF is ~ $10^7 / \left(1+(1.3*0.22)^2\right)^{1/2} \simeq 9.6*10^6$, which is close to the peak value shown in Figure 4.

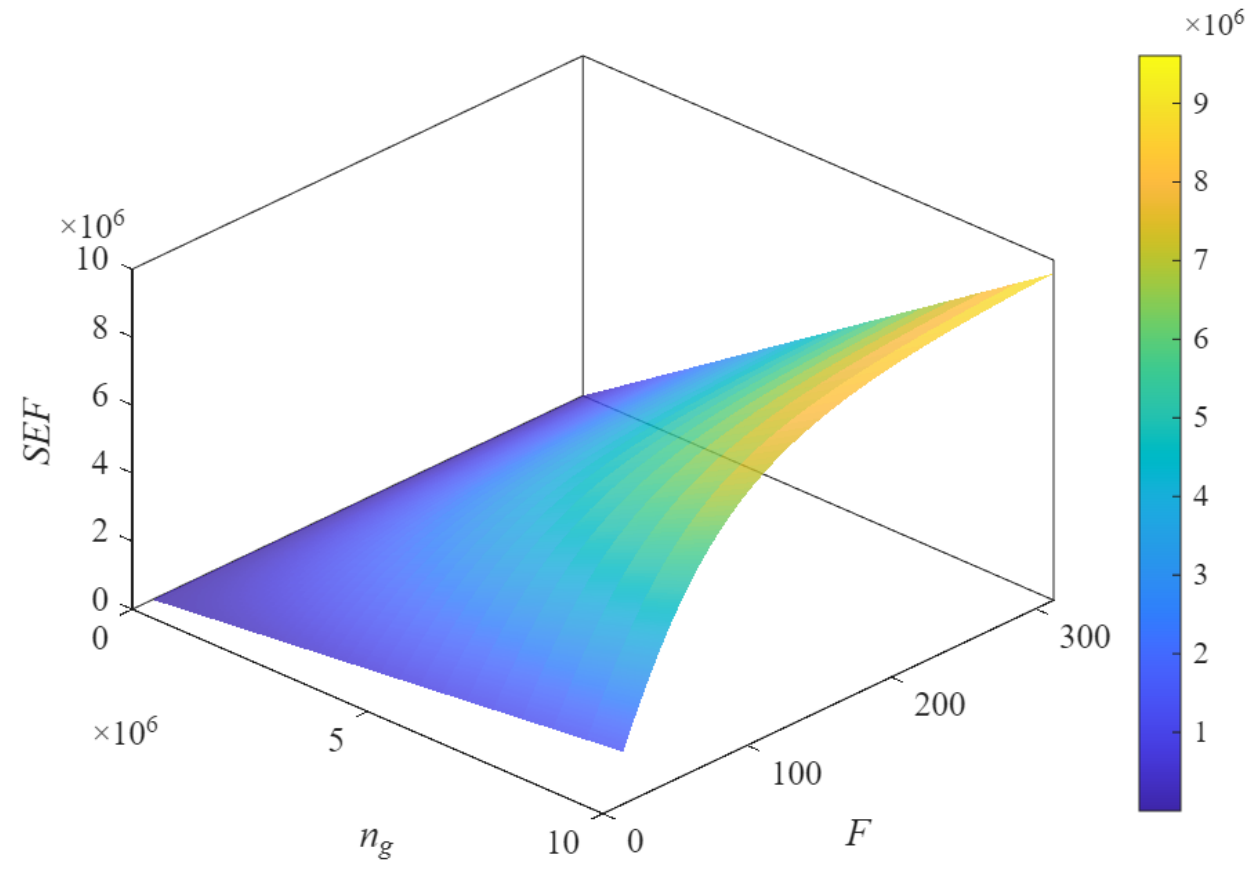

Figure 4. The SEF as a function of the group index and the finesse in the SLAFPC. The group index is varied from 1 to $10^7$ and the cavity finesse ranges from 1 to 313.

## 5. Effect of Absorption

Before we consider the effect of absorption along with all the other effects in the SLAFPC, it is helpful to look first at the case where there is no slow-light effect and the laser linewidth is a delta function. Using the same ring cavity we showed in Figure 1, we add a purely absorptive medium, ignoring any dispersion, uniformly to the cavity and express the absorption effect by a coefficient, $\alpha$. The laser field is attenuated by a factor of $e^{-\alpha l}$ after going through each leg. Therefore, in steady-state, we can write:

$$\begin{aligned} E_F^a &= tE_i + eE_i e^{jkl} e^{-\alpha l} r e^{j2kl} e^{-2\alpha l} r + tE_i \left(r^2 e^{jkL} e^{-\alpha L}\right)^2 + \cdots + tE_i \left(r^2 e^{jkL} e^{-\alpha L}\right)^n \\ E_o &= tE_F^b = tE_F^a e^{jkl} e^{-\alpha l} \end{aligned} \tag{41}$$

where $E_F^a$ and $E_F^b$ are as defined before and $L$=3$l$ is the length of the cavity. The transfer function then becomes:

$$H' = \left|\frac{E_o}{E_i}\right|^2 = \frac{T^2 e^{-2\alpha l}}{\left|1 - Re^{jkL} e^{-\alpha L}\right|^2} = \frac{T^2 e^{-2\alpha l}}{\left(1 - Re^{-\alpha L}\right)^2 + 4Re^{-\alpha L} \sin^2\left(kL/2\right)} \tag{42}$$

We define $\tilde{T} = Te^{-\alpha l} \equiv T\sigma$ and $\tilde{R} = Re^{-\alpha L} \equiv R\sigma$, where $\sigma \equiv \exp(-\alpha l)$ is the absorption factor for each arm. Then Eqn. (42) can be written as:

$$H' = \frac{\tilde{T}^2}{\left(1-\tilde{R}\right)^2 + 4\tilde{R}\sin^2\left(kL/2\right)} \equiv \frac{\tilde{a}_0}{1 + \tilde{b}_0 \sin^2\left(kL/2\right)} \tag{43}$$

where $\tilde{a}_0 = \tilde{T}^2 \Big/ \left(1-\tilde{R}\right)^2$ and $\tilde{b}_0 = 4\tilde{R} \Big/ \left(1-\tilde{R}\right)^2$. As before, we assume there is a resonance at $\omega_m$, which satisfies the condition $\sin\left[\omega_m L \big/ \left(2c_0\right)\right] = 1$, and denote the deviation from $\omega_m$ as $\tilde{\omega} = \omega - \omega_m$. Thus, around the frequency $\omega_m$, we can express Eqn. (43) as

$$H' \simeq \frac{\tilde{a}_0 \tilde{\tilde{\gamma}}_0^2}{\tilde{\tilde{\gamma}}_0^2 + \tilde{\omega}^2} \tag{44}$$

where $\tilde{\tilde{\gamma}}_0 = 2c_0 / L\sqrt{\tilde{b}_0}$. In the limit where there is no absorption, we get $\tilde{R} = R$ and $\tilde{T} = T$ so that Eqn. (4) is recovered.

It is apparent that Eqn. (44) has the same form as Eqn. (21), except for the definition of $\tilde{R}$ and $\tilde{T}$. Therefore, by comparing these two equations and Eqn. (21), we can explicitly write the absorption-modified transfer function when incorporating the slow-light effect and the laser linewidth at the same time:

$$H'_{SL} \simeq \frac{\tilde{a}\tilde{\tilde{\gamma}}_{EC}^2}{\tilde{\tilde{\gamma}}_{EC}^2 + \left(n_g \tilde{\omega}\right)^2} \tag{45}$$

where we have defined:

$$\tilde{\tilde{\gamma}}_{EC} = \frac{2c_0}{L} \cdot \frac{1}{\sqrt{\tilde{b}(\Delta)}}; \quad \tilde{a}(\Delta) = \frac{\tilde{T}}{\left(1-\tilde{R}^2\right)} \cdot \frac{\left(1-\Delta^2\tilde{R}^2\right)}{\left(1-\Delta\tilde{R}\right)^2}; \quad \tilde{b}(\Delta) = \frac{4\Delta\tilde{R}}{\left(1-\Delta\tilde{R}\right)^2} \tag{46}$$

Considering the limiting case where an ideal laser is used (i.e., $\Delta = 1$), we get $\tilde{a}(\Delta) = \tilde{a}_0$ and $b(\Delta) = \tilde{b}_0$ so that we recover Eqn. (44).

To determine the MMFS, we again assume that the output of the laser sent into the SLAFPC is expressed as $S_0 = N\tau_S$, where $N$ and $\tau_S$ are as defined before. Then the output signal for the SLAFPC becomes:

$$S = S_0 H'_{SL} = S_0 \cdot \frac{\tilde{a}\tilde{\tilde{\gamma}}^2_{EC}}{\tilde{\tilde{\gamma}}^2_{EC} + \tilde{\omega}^2} \tag{47}$$

The absorption only modifies the MMFS contributed from the vacuum mode. Therefore, the overall MMFS when taking into account absorption is found to be:

$$\Delta\tilde{\omega}_{SLAFPC,tot} = \sqrt{\Delta\omega^2_{SLAFPC,VM} + \Delta\omega^2_{SLAFPC,SE}} = \frac{\Gamma_0}{n_g}\sqrt{1+\left(\beta\tilde{\xi}\varepsilon_0\right)^2} \tag{48}$$

where we have defined:

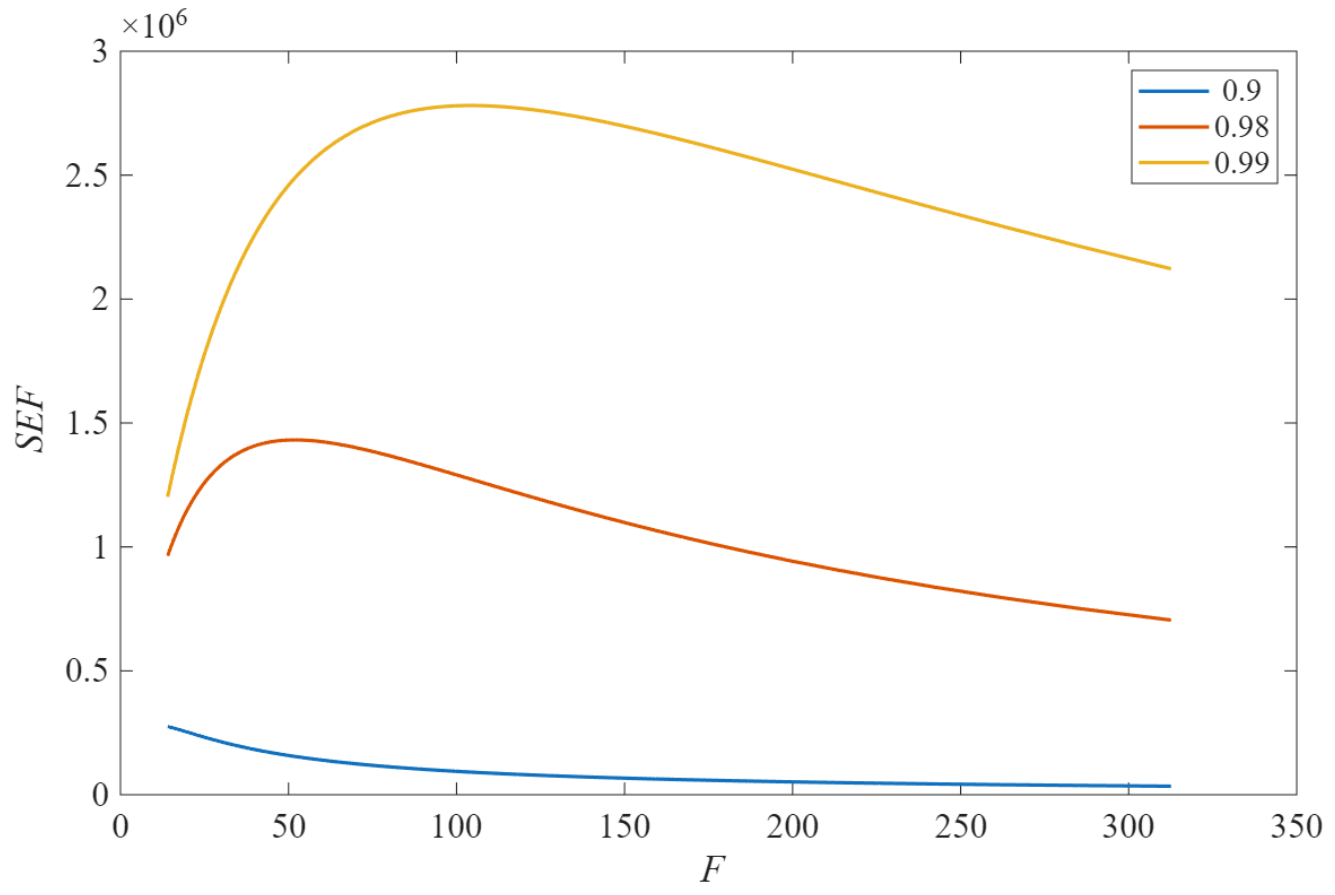

Figure 5. Plot of the SEF for the SLAFPC as a function of the cavity finesse, at different values of $\sigma$, the attenuation factor for each arm. The value of the group index used here is $10^7$.

$$\tilde{\xi} = e^{-\alpha\left(l-\frac{L}{2}\right)} \cdot \frac{\left(1-\tilde{R}\Delta\right)^2}{\tilde{T}^2}\sqrt{\frac{1-\tilde{R}^2}{\Delta\left(1-\Delta^2\tilde{R}^2\right)}} = e^{\alpha\frac{l}{2}} \cdot \frac{\left(1-\tilde{R}\Delta\right)^2}{\tilde{T}^2}\sqrt{\frac{1-\tilde{R}^2}{\Delta\left(1-\Delta^2\tilde{R}^2\right)}} \tag{49}$$

In the limit of no absorption ($\alpha = 0$), Eqn. (48) becomes the same as Eqn. (30), as expected.

The SEF in the presence of absorption becomes:

$$SEF = \frac{\Gamma_0}{\Delta\tilde{\omega}_{SLAFPC,tot}} = \frac{n_g}{\sqrt{1+\left(\beta\tilde{\xi}\varepsilon_0\right)^2}} \tag{50}$$

Figure 5 shows the SEF as a function of the finesse when $n_g = 10^7$, for three different values of $\sigma$, the attenuation factor for each arm. It can be seen from these plots that the inclusion of absorption significantly reduces the SEF. When the attenuation factor for each arm is close to unity (e.g., $\sigma = 0.99$), the value of the SEF increases rapidly at small values of the finesse but then diminishes with further increase of the finesse. For this degree of attenuation, which should be realizable, using, for example, coherent population trapping in cold atoms, the maximum SEF is found to be ~2.8*$10^6$. On the other hand, the SEF diminishes more rapidly when the absorption becomes more significant, as shown by the blue trace in Figure 5. Therefore, it is critical to keep the round trip loss as low as possible, in order to maximize the SEF achievable using a SLAFPC.

## 6. Conclusion

We have shown that the sensitivity of measuring the frequency shift of a laser can be significantly enhanced, compared to optimized standard techniques, by using a slow-light augmented Fabry-Perot Cavity (SLAFPC), in a manner analogous to what happens in a slow-light augmented Mach-Zehnder interferometer (SLAUMZI). This is due to the fact that an FPC is inherently unbalanced. We show how the degree of enhancement in sensitivity depends on the spectral width of the laser and the finesse of the FPC. We further find that the effect of the loss caused by the medium that produces the slow-light mechanism is very prominent for the SLAFPC and can be detrimental for sensitivity enhancement. For potentially realizable conditions, we show that an SEF of ~2.8*$10^6$ can be achieved using a SLAFPC.

**Supplement: Effect of Laser Linewidth on the FPC Transfer Function**

As shown in the main text, the transfer function for the FPC, when taking into account the STL, can be expressed as:

$$H = T^2 \sum_{n=0}^{\infty}\sum_{m=0}^{\infty} \left\langle \varepsilon_n \varepsilon_m^* \right\rangle = T^2 \left( \sum_{n=0}^{\infty} \left|\varepsilon_n\right|^2 + \sum_{\substack{n=0 \\ n\neq m}}^{\infty}\sum_{m=0}^{\infty} \varepsilon_n \varepsilon_m^* e^{-Q|m-n|} \right) \equiv T^2 \left(A+B\right) \tag{S.1}$$

For simplicity, we define $\Delta = e^{-Q}$ and $\tilde{\delta} = kL$, and explicitly write $A$ and $B$ in terms of $\Delta$ and $\tilde{\delta}$:

$$A = \sum_{n=0}^{\infty} \left|\varepsilon_n\right|^2 = \sum_{n=0}^{\infty} \left| R^n e^{jn\tilde{\delta}} \right|^2 = \sum_{n=0}^{\infty} R^{2n} = \frac{1}{1-R^2} \tag{S.2}$$

$$B = \sum_{n=0}^{\infty}\sum_{m=0}^{\infty} \varepsilon_n \varepsilon_m^* e^{-Q|m-n|} = \sum_{n=0}^{\infty}\sum_{m=0}^{\infty} \left(R^n e^{jn\tilde{\delta}}\right)\left(R^m e^{jm\tilde{\delta}}\right)^* \Delta^{|m-n|} \tag{S.3}$$

To evaluate $B$, we break it up into an infinite number of groups, each designated as Group($N$), where the value of $N$ ranges from 1 to infinity: $N = 1,2,3....\infty$. Each Group($N$) represents two sets of terms: one [designated as Group($N_A$)] in which the value of $n$ is held fixed at ($N$-1), while the values of $m$ range from $N$ to infinity, and another [designated as Group($N_B$)] in which the value of $m$ is held fixed at ($N$-1), while the values of $n$ range from $N$ to infinity. We now consider specifically a few of these terms in order to see the general pattern.

Group(1):

$$\text{Group}\left(1_A\right): \begin{cases} n=0, m=1 \rightarrow \varepsilon_0 \varepsilon_1^* \Delta^1 = R^1 \Delta^1 e^{-j\tilde{\delta}} \\ n=0, m=2 \rightarrow \varepsilon_0 \varepsilon_2^* \Delta^2 = R^2 \Delta^2 e^{-j2\tilde{\delta}} \\ n=0, m=3 \rightarrow \varepsilon_0 \varepsilon_3^* \Delta^3 = R^3 \Delta^3 e^{-j3\tilde{\delta}} \\ \vdots \end{cases} \tag{S.4}$$

$$\text{Group}\left(1_B\right): \begin{cases} m=0, n=1 \rightarrow \varepsilon_1 \varepsilon_0^* \Delta^1 = R^1 \Delta^1 e^{j\tilde{\delta}} \\ m=0, n=2 \rightarrow \varepsilon_2 \varepsilon_0^* \Delta^2 = R^2 \Delta^2 e^{j2\tilde{\delta}} \\ m=0, n=3 \rightarrow \varepsilon_3 \varepsilon_0^* \Delta^3 = R^3 \Delta^3 e^{j3\tilde{\delta}} \\ \vdots \end{cases} \tag{S.5}$$

Thus, the Group(1) type terms would add up to yield:

$$B_{Group(1)} = \sum_{n=-\infty}^{+\infty} \left(R\Delta\right)^{|n|} e^{jn\tilde{\delta}} - \left(R\Delta\right)^{|n|} e^{jn\tilde{\delta}}\Big|_{n=0} = \sum_{n=-\infty}^{+\infty} \left(R\Delta\right)^{|n|} e^{jn\tilde{\delta}} - 1 \tag{S.6}$$

Group(2):

$$
\text{Group}(2_A):\begin{cases} n=1, m=2 \to \varepsilon_1\varepsilon_2^*\Delta^1 = R^3\Delta^1 e^{-j\tilde{\delta}} \\ n=1, m=3 \to \varepsilon_1\varepsilon_3^*\Delta^2 = R^4\Delta^2 e^{-j2\tilde{\delta}} \\ n=1, m=4 \to \varepsilon_1\varepsilon_4^*\Delta^3 = R^5\Delta^3 e^{-j3\tilde{\delta}} \\ \vdots \end{cases} \tag{S.7}
$$

$$
\text{Group}(2_B):\begin{cases} m=1, n=2 \to \varepsilon_2\varepsilon_1^*\Delta^1 = R^3\Delta^1 e^{j\tilde{\delta}} \\ m=1, n=3 \to \varepsilon_3\varepsilon_1^*\Delta^2 = R^4\Delta^2 e^{j2\tilde{\delta}} \\ m=1, n=4 \to \varepsilon_4\varepsilon_1^*\Delta^3 = R^5\Delta^3 e^{j3\tilde{\delta}} \\ \vdots \end{cases} \tag{S.8}
$$

Thus, the Group(2) type terms would add up to yield:

$$
B_{Group(2)} = R^2\left(\sum_{n=-\infty}^{+\infty}(R\Delta)^{|n|} e^{jn\tilde{\delta}} - 1\right) \tag{S.9}
$$

Group(3):

$$
\text{Group}(3_A):\begin{cases} n=2, m=3 \to \varepsilon_2\varepsilon_3^*\Delta^1 = R^5\Delta^1 e^{-j\tilde{\delta}} \\ n=2, m=4 \to \varepsilon_2\varepsilon_4^*\Delta^2 = R^6\Delta^2 e^{-j2\tilde{\delta}} \\ n=2, m=5 \to \varepsilon_2\varepsilon_5^*\Delta^3 = R^7\Delta^3 e^{-j3\tilde{\delta}} \\ \vdots \end{cases} \tag{S.10}
$$

$$
\text{Group}(3_B):\begin{cases} m=2, n=3 \to \varepsilon_3\varepsilon_2^*\Delta^1 = R^5\Delta^1 e^{j\tilde{\delta}} \\ m=2, n=4 \to \varepsilon_4\varepsilon_2^*\Delta^2 = R^6\Delta^2 e^{j2\tilde{\delta}} \\ m=2, n=5 \to \varepsilon_5\varepsilon_2^*\Delta^3 = R^7\Delta^3 e^{j3\tilde{\delta}} \\ \vdots \end{cases} \tag{S.11}
$$

Thus, the Group(3) type terms would add up to yield:

$$
B_{Group(3)} = R^4\left(\sum_{n=-\infty}^{+\infty}(R\Delta)^{|n|} e^{jn\tilde{\delta}} - 1\right) \tag{S.12}
$$

It is now clear that the Group(N) type terms would add up to yield:

$$
B_{Group(N)} = R^{2(N-1)}\left(\sum_{n=-\infty}^{+\infty}(R\Delta)^{|n|} e^{jn\tilde{\delta}} - 1\right) \tag{S.13}
$$

Summing over the contribution of the infinite number of group, we then get:

$$
B = \sum_{M=0}^{\infty} R^{2M}\left(\sum_{n=-\infty}^{+\infty}(R\Delta)^{|n|} e^{jn\tilde{\delta}} - 1\right) = \frac{1}{1-R^2}\left(\sum_{n=-\infty}^{+\infty}(R\Delta)^{|n|} e^{jn\tilde{\delta}} - 1\right) \tag{S.14}
$$

Noting that $A = 1/(1-R^2)$ and $T = (1-R)$, we get:

$$H\left(\tilde{\delta}\right)=\frac{\left(1-R\right)^2}{1-R^2}\sum_{n=-\infty}^{+\infty}\left(R\Delta\right)^{|n|}e^{jn\tilde{\delta}} \tag{S.15}$$

To evaluate the sum, we write:

$$\begin{aligned}S\left(\delta\right)&\equiv\sum_{n=-\infty}^{+\infty}\left(R\Delta\right)^{|n|}e^{jn\tilde{\delta}}=\sum_{n=0}^{+\infty}\left(R\Delta\right)^{n}e^{jn\tilde{\delta}}+\sum_{n=0}^{-\infty}\left(R\Delta\right)^{|n|}e^{jn\tilde{\delta}}-1\\&=\sum_{n=0}^{+\infty}\left(R\Delta\right)^{n}e^{jn\tilde{\delta}}+\sum_{n=0}^{-\infty}\left(R\Delta\right)^{-n}e^{-j(-n)\tilde{\delta}}-1=\frac{1-R^2\Delta^2}{\left(1-R\Delta\right)^2+4R\Delta\sin^2\left(\tilde{\delta}/2\right)}\end{aligned} \tag{S.16}$$

Plugging into Eqn.(S.14) and substituting $\tilde{\delta}=kL=\omega L/c_0$ , we get:

$$H\left(\delta,\Delta\right)=\frac{a\left(\Delta\right)}{1+b\left(\Delta\right)\sin^2\left(\omega L/2c_0\right)} \tag{S.17}$$

where

$$a\left(\Delta\right)=\frac{\left(1-R\right)^2}{1-R^2}\cdot\frac{1-\Delta^2R^2}{\left(1-\Delta R\right)^2}=\frac{\left(1-R\right)}{1+R}\cdot\frac{1-\Delta^2R^2}{\left(1-\Delta R\right)^2} \tag{S.18}$$

$$b\left(\Delta\right)=\frac{4\Delta R}{\left(1-\Delta R\right)^2} \tag{S.19}$$

which agrees with the results shown in Ref. [30] in the limit of no loss.